\documentclass[a4paper]{svjour3}

\usepackage{graphicx}
\usepackage{times}
\usepackage{soul}
\usepackage[utf8]{inputenc}
\usepackage{url}
\usepackage{amsmath}
\usepackage{amsfonts}
\usepackage{amssymb}
\usepackage{comment}
\usepackage{soul}
\usepackage{fancyvrb}
\usepackage{relsize}
\usepackage{booktabs}
\usepackage[hybrid]{markdown}

\usepackage[usenames,dvipsnames]{xcolor}

\usepackage[textsize=tiny,textwidth=2cm]{todonotes}
\newcommand{\snote}[1]{\todo[color=CornflowerBlue,fancyline]{SC: #1}}
\newcommand{\nnote}[1]{\todo[color=LimeGreen,fancyline]{NO: #1}}

\newcommand{\V}{\mathcal{V}} 
\newcommand{\Dv}{{\mathcal{D}_v}} 
\newcommand{\ES}{S} 
\newcommand{\BS}{B} 
\newcommand{\Act}{A} 
\newcommand{\Strat}{Strat} 

\title{Designing for Accountable Agents: a Viewpoint
}
                         
\author{Stephen Cranefield \and Nir Oren }

\institute{S. Cranefield (Corresponding author)
          \at School of Computing, University of Otago, Dunedin, New Zealand, \\ \email{stephen.cranefield@otago.ac.nz}
          \and
          N. Oren
          \at Computing Science, University of Aberdeen, Aberdeen, UK, \\ \email{n.oren@abdn.ac.uk} }

\def\makeheadbox{\relax}
\makeatletter
\let\@date\null
\makeatother

\begin{document}

\maketitle


\begin{abstract}
AI systems are becoming increasingly complex, ubiquitous and autonomous, leading to increasing concerns about their impacts on individuals and society. In response, researchers have begun investigating how to ensure that the methods underlying AI decision-making are transparent and their decisions are explainable to people and conformant to human values and ethical principles. As part of this research thrust, the need for accountability within AI systems has been noted, but this notion has proven elusive to define; we aim to address this issue in the current paper. Unlike much recent work, we do not address accountability within the human organisational processes of developing and deploying AI; rather we consider what it would it mean for the agents within a multi-agent system (MAS), potentially including human agents, to be accountable to other agents or to have others accountable to them. 

In this work, we make the following contributions: we provide an in-depth survey of existing work on accountability in multiple disciplines, seeking to identify a coherent definition of the concept;
we give a realistic example of a multi-agent system application domain that illustrates the benefits of enabling agents to follow accountability processes, and we identify a set of research challenges for the MAS community in building accountable agents, sketching out some initial solutions to these, thereby laying out a road-map for future research. Our focus is on laying the groundwork to enable autonomous elements within open socio-technical systems to take part in accountability processes.

\end{abstract}

\section{Introduction}

Accountability is increasingly invoked whenever decisions---human or algorithmic---produce significant social, economic, or ethical consequences. Governments, public bodies, corporations, and individuals are routinely expected to give an account of their actions, justify their decisions, and accept the consequences when things go wrong. Although accountability can support sanctions and blame assignment, its broader purpose is to promote learning, trust, transparency, and improved future behaviour. For this reason, accountability has long been treated as a central concept across law, philosophy, governance studies, and the social sciences.

Despite this prominence, accountability remains conceptually unsettled. Across disciplines, the term has been described as “expansive, ambiguous and often enigmatic” \cite{AccountabilityasaCulturalKeyword}, “slippery” \cite{AccountabilitiesCh1}, and capable of meaning “all things to all people” \cite{FoxTransparencyAndAccountability}. Even within a single domain, definitional divergence is growing: Kaba’s review of 217 articles related to the accountability of non-governmental organisations identified 113 distinct conceptualisations of accountability, with the number of competing definitions increasing sharply over time \cite{NGO_accountability_2021}. This conceptual plurality is not simply an academic curiosity---it challenges any attempt to design systems or agents that must operate within accountability relationships.

At the same time, artificial intelligence systems---often operating autonomously, and increasingly deployed in safety‑critical or socially consequential settings---are prompting growing demands for ethical AI, transparency, and regulatory oversight \cite{jobin19global}. Current research in computer science, including within the multi-agent systems community, often responds to these demands by asking how to \emph{engineer} accountability: building standards, architectures, and processes that constrain or document system behaviour, enable post‑hoc explanations, or determine responsibility for failures. While this line of work is vital, it views AI artefacts primarily as objects of accountability: entities whose behaviour must be tracked, justified, or constrained by human or institutional mechanisms.

This paper proposes a different perspective. We ask: What would it mean for an autonomous software agent---not merely to be held accountable---but to act as a participant in accountability relationships? That is, how could an agent:

\begin{enumerate}
\item Identify and implement the processes required to meet its accountability obligations in open, dynamic environments; and
\item Engage with other agents---human or artificial---by giving accounts, demanding accounts, and using the accountability process to improve its own functioning?
\end{enumerate}

Where existing research focuses on system‑level mechanisms for enforcing accountability, our interest lies in the capabilities an individual agent requires to reason about, undertake, and benefit from accountability interactions in an open system. This distinction mirrors the classic difference between closed, centrally designed multi‑agent systems and open environments in which autonomous entities must independently interpret norms, expectations, and social processes.

A major obstacle to developing such agents is the limited engagement of the multi‑agent systems community with the rich conceptual and empirical work on accountability in other fields. As a result, MAS approaches often rely on an implicit, intuitive notion of accountability---often without defining it at all---despite substantial evidence that no shared definition exists across disciplines. Before accountable agency can be meaningfully explored, the MAS field requires a clearer synthesis of the purposes, structures, and processes that constitute accountability in human organisations.

In this paper, we therefore (i) review key literatures on accountability across multiple domains; (ii) identify conceptual features and common patterns that matter for autonomous agents; and (iii) outline a research agenda for accountable agency in open multi‑agent systems. Crucially, while past MAS work has emphasised responsibility determination (e.g., blame assignment for failures \cite{halpern2018towards,MuNajibOren25,ParkerGL23}), we focus on the stages before and after responsibility attribution---especially the creation, delivery, and interpretation of accounts, and the role these processes play in shaping future behaviour.

This paper significantly extends our earlier work \cite{CranOrenVasc_2019} by offering a deeper synthesis of the cross‑disciplinary literature, clarifying the notion of accountable agency, and outlining the research challenges that must be addressed. Our aim is not to propose complete solutions, but to articulate a conceptual agenda and motivate future research.

The remainder of this paper is organised as follows. Section \ref{sec:what_is_accountability} synthesises existing definitions of accountability and its purposes. Section \ref{sec:scenario} presents a MAS scenario illustrating how accountability interactions can improve system behaviour. In Section \ref{sec:challenges} we outline the research challenges in developing accountable autonomous agents. Section \ref{sec:conclusions} concludes.

\section{What is accountability and why do we need it?}
\label{sec:what_is_accountability}

One of the goals of this paper is to demonstrate why accountability is a useful concept for software agents. In Section \ref{sec:purpose}  considering the purposes of accountability in human systems, as discussed in the literature. 
%
Our goal is to seek clarity on the purpose of accountability, independently of how it might be realised in different settings, or whether or not it might be considered an ethical imperative. Following this, in Section \ref{sec:conceptions}, we consider how accountability has been conceptualised by researchers in other fields. Accountability is closely related to various other concepts, including responsibility, answerability and liability, and we differentiate these notions in Section \ref{sec:related_concepts}. We then consider how multi-agent researchers have considered and utilised accountability. 


\subsection{The purpose of accountability}\label{sec:purpose}

When setting out to design accountable software agents it is important to consider the functional purpose of accountability, that is, whether accountability is simply a feature that satisfies a human desire to feel empowered, or if there are system-level benefits. In the former case, there may be no point in creating accountable agents unless they are interacting with people. In the latter case, it is necessary to identify the benefits that we wish our agents (or their society) to enjoy.

Several authors have argued that the main purpose of accountability is to provide some form of control. Mulgan~\cite{doi:10.1111/1467-9299.00218} notes that:
\begin{quote}
  ``The core sense of accountability is clearly grounded in the general purpose of making agents or sub-ordinates act in accordance with the wishes of their superiors. Subordinates are called to account and, if necessary, penalized as means of bringing them under control.''
\end{quote}

Lindberg \cite{Lindberg2009} makes a similar point, arguing that ``accountability belongs to a class of concepts under the more general category of `methods of limiting power'\hspace{0.25ex}''.

An examination of accountability in the public policy context is provided by Bovens~\cite{doi:10.1111/j.1468-0386.2007.00378.x}, who also highlights the aspect of control, as well as two other purposes of accountability, which we discuss below:
\begin{quote}
``So why is accountability important? \dots\ In the academic literature and in policy publications about public accountability, three answers recur, albeit implicitly, time and again. Accountability is important to provide a democratic means to monitor and control
government conduct, for preventing the development of concentrations of power, and
to enhance the learning capacity and effectiveness of public administration.''
\end{quote}

In the context of multi-agent systems, the monitoring and control aspect of accountability provides motivation for agents to perform well: a rational agent (as software agents are generally designed to be) will be likely to prioritise goals for which it is accountable, and devote more resources to them so as to avoid being sanctioned in the event of sub-standard performance or failure. At the same time, accountability also provides the agent with the ability to point out mitigating circumstances if failure occurs, allowing it to avoid the impacts of such failure due to circumstances outside its control.

We consider that the second aspect of Bovens's analysis quoted above (avoiding concentration of power) is probably less relevant to multi-agent systems than to human society; however, we consider the third purpose (enhancing the learning capacity and effectiveness of the wider system) to be important. Bovens elaborates on this as follows:
\begin{quote}
``The purpose of public accountability is to induce the executive branch to learn. The possibility of sanctions from clients and other stakeholders in their environment in the event of errors and shortcomings motivates them to search for more intelligent ways of organising their business. Moreover, the public nature of the accountability process teaches others in similar positions what is expected of them, what works and what does not.''
\end{quote}
The last sentence implies a norm-alignment and knowledge-spreading function of accountability. As Bovens notes elsewhere in his article, ``Norms are (re)produced, internalised and, where necessary, adjusted through accountability'' \cite{doi:10.1111/j.1468-0386.2007.00378.x}.

Day and Klein \cite{AccountabilitiesCh1} highlight the role of accountability in setting professional standards that help protect the public from incompetent or otherwise unfit practitioners. This is a consequence of the rise of \emph{professional accountability} in the 20th century, whereby professionals are accountable (only) to peers within professional associations. These associations are ``granted monopoly rights of practice in return for policing the competence of their members'' \cite{AccountabilitiesCh1}. In critical fields such as medicine, these may be backed up by national tribunals that are constituted by legislation giving them explicit goals related to accountability within the profession.


Mulgan \cite{doi:10.1111/1467-9299.00218} summarises the analysis of political accountability by March and Olsen \cite{march_democratic_1995} in which ``explanation and justification [lie] at the core of accountability'' and ``[c]alling people to account means inviting them to explain and justify their actions within two competing logics, that of consequences and that of appropriateness''. Thus, in this view, a key purpose of accountability is to require explanations of actions that led to poor outcomes, or where correct processes were not followed.

Greiling \cite{greiling2014AccountabilityTrust} highlights a long strand of research finding that accountability can affect the level of trust placed in an entity. For example, ``distrust in government \dots\ leads to intensive public accountability obligations aiming at enabling citizen trust in the government'', with such controls facilitating trust if they are ``rule-based, impersonal, and embedded in the structures, norms and policies of a public agency'' \cite{choudhury2008}. The control provided by accountability, and concomitant process improvement, leads to higher performance of an organisation, with this performance leading to increased trust \cite{halachmi2010}. By serving as a signalling mechanism, accountability signals competence, also boosting trust \cite{greiling2014AccountabilityTrust}.

Given the research cited above, the purpose of accountability in human interactions appear to be about control. In this context, account-givers or \emph{accountors} are prevented from acting in an undesirable manner, or are required to explain why they should not be punished. Accountability also facilitates the spreading of knowledge and system improvement, for example by having the account-taker or \emph{accountee} specify new ways of working, leading to increased trust over time.
%
Many of the goals of accountability in human systems are also applicable to accountability-aware multi-agent systems. Within an MAS, accountability can provide control through the potential for sanctions imposed by an accountee, and may constrain agent behaviour as the accountor is aware that they may be held to account. The ability of the accountee to require corrective behaviour and to recommend changes in operating processes and protocols facilitates system improvement, while the dialogical element of accountability (discussed in the next section) can support the assignment of blame or responsibility. This latter element provides another benefit, namely that by requiring an explanation, (complex inscrutable) software agents must be able to explain their behaviour to an accountee, providing transparency and explainability. The benefits of such scrutiny have been highlighted within the explainable AI (XAI) literature \cite{miller2019}.

March and Olsen \cite{march_democratic_1995} have highlighted several downsides to accountability, noting that when aware that they may be held to account, humans pay increasing attention to extraneous factors, to ``increase caution about change and to reduce risk-taking''. These downsides appear to originate from human psychological underpinnings. Therefore, artificial systems hardwired to demonstrate accountability  may cause similar issues to arise due to human biases encoded into their design. In contrast such issues should not occur in rational autonomous systems designed to be able to reason about accountability.

While the discussion above is informed by research on accountability within human social systems, we argue that many of its purposes
also apply to 
(software) multi-agent systems, especially those in which people and software agents interact. For example, accountability has a role to play in motivating good performance, and in monitoring and control (when one agent or group is a subordinate of another). It can also allow for incremental system improvement through learning or instruction, e.g.,~one agent may send new plans to another agent as an outcome of an accountability dialogue, and can enable the alignment and spreading of norms. When human users or partners are involved, we also see accountability contributing to the alignment of values.

\subsection{Conceptions of accountability}\label{sec:conceptions}

We previously noted that many definitions of accountability exist and that a broad understanding of the concept is important for its application. In this section we survey the literature on accountability from multiple fields, including policy-making, sociology, management, and computer science. Our aim is to identify the key requirements that an autonomous agent would need to satisfy to be considered accountable.

Day and Klein provide an in-depth analysis of \emph{public accountability} \cite{AccountabilitiesCh1}. They discuss the origins and evolution of accountability as a mechanism to foster both political and financial responsibility by government officials. They describe the earliest known processes of \emph{political accountability} in ancient Athens, where elected officials were motivated to govern responsibly through an obligation to justify their conduct in a public assembly ten times a year and win a vote of confidence. Failure could lead to a trial by jury, with potential outcomes including death and exile. In contrast, Day and Klein argue that \emph{managerial accountability} has distinct roots in the practice of auditing the accounts of an estate to ensure financial responsibility by a steward given authority to manage its affairs. As far back as the Greek city states and Egyptian kingdoms, the role of a state auditor existed to verify, according to clear standards, that true and accurate accounts had been kept. This is in contrast to the emphasis on ``pursuasive accounts'' in political accountability, ``where there may be disagreement about \dots\ the criteria for judging actions or policies to be right or justified'' \cite{AccountabilitiesCh1}.

Schedler \cite{andreas_schedler_conceptualizing_1999} also notes the ``etymological ambivalence of accountability'' due to two different meanings of ``accounts'': financial accounts containing ``detailed information prepared according to certain standards of classification and accuracy'' and narrative accounts, which are ``legitimating tales''. The latter are particularly relevant to the area of political accountability, where they take the form of explanations that ``excuse or justify questionable behavior by proposing a normative status for the behavior'' and ``convey implicit or explicit information about actors’ roles, methods, and contexts of action''.

Bovens et al.~\cite{PublicAccountability} discuss the views of accountability in the social psychology, accounting, public administration, political science, international relations and constitutional law literature. In particular, they characterise the definition of accountability in the social psychology literature as the expectation that one may be asked, often by an authority or one's superior, to justify one's thoughts, beliefs, or actions'', and note that this field ``primarily focuses on the communicative interaction between an agent and an audience and its effects on his (or her) choices and behavior''.

From these analyses, our first observation is that accountability revolves around an \emph{accountability relationship} between an accountor (who provides the account) and accountee (who requests, observes, or judges the account given).

Bovens et al.\@ observe that there is a ``minimal consensus'' in the academic literature on the nature of accountability. Schillemans~\cite{SchillemansWorkingPaper} expresses this consensus in the following way:

\begin{enumerate}
  \item Accountability is about answerability (towards those with a legitimate claim in some agents' work);
  \item Accountability is a relational concept ascribed to agents performing tasks for others;
  \item Accountability is retrospective; and
  \item Accountability is behaviour-oriented, concerning itself with performance and results in spheres including financial management, regularity and normative and professional standards.
\end{enumerate}

\label{schillemans_phases}
Furthermore, Schillemans notes that accountability consists of ``three analytically distinct phases''.
\begin{enumerate}
  \item During the \emph{information phase}, the accountor renders an account on [their] conduct and performance to a significant other.
  \item In the \emph{debating phase} the accountee assesses the \ldots\  transmitted information. Additional discussion may take place, with the accountee asking for additional information and judging the accountor, while the latter answers questions and justifies and defends their course of action.
  \item In the \emph{sanctions} or \emph{judgment phase}, the accountee reaches a concluding judgment and decides whether/how to sanction the accountor.
\end{enumerate}

Schillemans' focus is on what occurs when an accountor is held to account. We note in passing that prior to this---and particularly in the context of artificial agents---an accountor may need to decide how to act given an awareness that they may be called to give an account, and may also need to decide what information to preserve. An accountee must decide when the rendering of an account may be required.


\label{sec:remedies}
While the analysis of Schillemans focuses on sanctioning as a possible outcome of the accountability process, Gibbons's discussion of \emph{social accountability} \cite{gibbonssocialaccountability2014} shows that there are other possible outcomes, e.g., there are strong expectations from the public that \emph{remedies} will be provided for failures in service delivery. Gibbons discusses four types of remedy. The most common is \emph{redress}: ensuring the intended level of rights fulfilment is achieved by correcting the error or through alternative means. Other short-term remedies are \emph{sanctioning} responsible parties and providing monetary \emph{compensation} for harm caused. A longer-term remedy is to \emph{guarantee non-repetition} by improving legislation, planning and budgetary processes and training, implementing new monitoring and enforcement mechanisms, and ensuring full and public disclosure of the truth.

Emanuel and Emanuel \cite{AccountabilityHealthCare} give a definition of accountability in the domain of healthcare: ``Accountability \dots\ entails procedures and processes by which one party provides a justification and is held responsible for its actions by another party that has an interest in the actions''. They consider the following components of accountability: the \emph{locus} of accountability, i.e.~who can be held accountable, the \emph{domain} of accountability, i.e.~the ``activity, practice, or issue for which a party can legitimately be held responsible and called on to justify or change its action''
and the \emph{procedures} of accountability, divided into evaluation of compliance and dissemination of evaluations to seek ``responses or justifications'' from accountable parties. 

Lindberg \cite{Lindberg2009} focuses on political and financial notions of accountability, and notes that accountability provides a mechanism---following the transfer of some power for decision-making from accountee to accountor---for the accountee to require the accountor to justify their actions and, if necessary, to sanction the accountor (including removing the power transferred to them). Lindberg notes several fundamental characteristics that (they claim) are present in any type of accountability relation.
\begin{enumerate}
    \item The actors taking part in the relationship (i.e., the accountor and accountee).
    \item The area, responsibility or domain which is the subject of the account (i.e., the relevant power given to the accountor).
    \item The right of the accountee to ask the accountor to justify their decisions.
    \item The right of the accountee to sanction the accountor if the latter fails to justify their decisions. 
\end{enumerate}
Depending on what the source of the accountee's powers are, their strength, and the relationship between accountor and accountee, Lindberg identifies 12 different subtypes of accountability.

Broadly, all of these conceptions of accountability identify a (group of) accountors and accountees; a domain whereby the accountor can be held to account; a recognition that the accountor has obligations placed on them; the process by which the accountor is held to account; and the powers held by the accountee to impose sanctions, demand changes and provide redress for harms once the accountability process is concluded.

\subsection{Related concepts}
\label{sec:related_concepts}

There is an abundance of concepts discussed in the literature as synonyms or dimensions of accountability. In this section we give an overview of this richness of vocabulary, and make some attempts to relate different authors' conceptions. However, it is not our aim to propose a single typology of accountability-related concepts.

\label{sec:moral_forces}
Dubnick~\cite[Fig.~2.4]{AccountabilityasaCulturalKeyword} categorises various concepts related to accountability that are motivated by moral forces \cite[p.401]{noznick81}:
Such external forces, i.e., \emph{moral pulls}, lead to concepts such as liability, answerability, responsibility and responsiveness (which are terms used, respectively, in the legal, organisational, professional and political settings). 
In contrast, \emph{moral pushes} arise due to intrinsic motivations, and are related to concepts such as obligation, obedience, fidelity and amenability (in the same four settings, respectively).

Elsewhere, Dubnick analyses ``four general types of mechanisms that demand account-giving responses: answerability, blameworthiness, liability and attributability''~\cite{dubnickAccountabilityEthicsReconsidering2003}, and considers how each of these relates to ethical decision-making. These mechanisms can be viewed as specifying elements or social structures of the domain under which an accountability relationship can form. 

\emph{Answerability}
is defined by Dubnick as the expectation ``to respond to calls for giving an account upon demand''. 
 Dubnick positions answerability as promoting ``ethical mechanisms stressing the consequences of their actions vis-\`{a}-vis the expectations of those to whom they are answerable''. However, we also see this as an important step to initiate a dialogue that may lead to system improvement.

Dubnick's concept of \emph{blameworthiness} relates to a person being held to account not because of a specific task or responsibilities, but rather due to their social or organisational status (e.g.,~as a military commander); however, other definitions of blameworthiness appear in the literature, as discussed under the topic of responsibility below. Dubnick argues that blameworthiness has promoted the ``establishment of credibility for public administrators as autonomous ethical actors''.

\emph{Liability} corresponds to a legalistic view, in which ``actions are guided and assessed according to rules that carry sanctions for non-compliance'', irrespective of one's organisational role or status. According to Dubnick, this fosters ethical reasoning based on ``procedural requirements, legal standards or precedent''. 

Finally, \emph{attributability} is when a worker is held to account for non-workplace behaviour that is ``attributed'' to him or her, leading to questions about his or her character. This may lead to an ethical approach based around notions of citizenship, virtue and character.

\label{sec:koppell_responsiveness}
In the context of public administration, Koppell  \cite{koppellPathologiesAccountabilityICANN2005} discusses five dimensions of accountability: transparency, liability, controllability, responsibility and responsiveness. \emph{Transparency} requires an accountable party to ``explain or account for its actions'', and appears similar to Dubnick's concept of answerability. Transparency also has other definitions in the literature, discussed below. 
Koppell defines \emph{liability} as the requirement for ``individuals and organizations to face consequences that are attached to performance''. \emph{Controllability} is the ability of one party to induce the behaviour of the other, making them accountable in the process. \emph{Responsibility} relates to fidelity to law and behavioural norms (this and other definitions are discussed below). Finally, \emph{responsiveness} refers to an organisation's attention to the demands and needs of the people being served. Here, \emph{demands} refers to preferences elicited through polling or consultation, whereas \emph{needs} refers to the goals the organisation ``ought to be pursuing'' to serve their constituents. 

\label{sec:fox_transparency}
Fox~\cite{FoxTransparencyAndAccountability} discusses the relationship between transparency and accountability in human institutions, which (he claims) is conventionally expressed as ``transparency generates accountability''. After reviewing the empirical literature he concludes that neither of these concepts necessarily generates the other. Their presence depends on which of the following three institutional capacities are present: 
to disseminate and provide and access to information, to demand explanations (answerability) and to sanction and/or compensate. The first of these on its own leads to an ``opaque'' form of transparency, where existing information is made available. Adding answerability allows new information about institutional behaviour to become available. This leads to full (\emph{clear}) transparency, but also creates a \emph{soft} form of accountability. Fox considers that ``this \emph{capacity} to produce answers permits the construction of the \emph{right} to accountability''. However, the full \emph{hard} type of accountability is only present when the ability to sanction and compensate is added. 


The relationships between accountability and responsibility seem especially subject to varying viewpoints. Dubnick~\cite{AccountabilityasaCulturalKeyword} notes that one can be ``responsible for some event, for example the marriage of two people who met because (one) did not take the empty seat between them on the bus, without being held to account for it''. Eshleman~\cite{sep-moral-responsibility} discusses various philosophical views on \emph{moral} responsibility. The \emph{accountability} view holds that ``an agent is responsible, if and only if it is appropriate for us to hold her responsible, or accountable, via the reactive attitudes \dots\ (e.g.~resentment)''.  Another influential view, referred to by Eshleman as the \emph{answerability} view, is that ``someone is responsible for an action or attitude just in case it is connected to her capacity for evaluative judgment in a way that opens her up, in principle, to demands for justification from others''. Both van de Poel~\cite{vandePoel2011} and Davis~\cite{DBLP:journals/see/Davis12} highlight the diversity of interpretations of responsibility, giving (distinct) lists of nine senses of the term.

In the practice of business management, a Responsible, Accountable, Consulted, and Informed (RACI) matrix is a recognised tool to map where responsibility and accountability are assigned for activities~\cite{PMBOK_Guide}. In this context, the responsible parties are those who work on the activity (responsibility may be shared), whereas the accountable party is the (unique) person with ``yes or no authority'' over the activity and ``about whom it is said `The buck stops here'\,''~\cite{BusinessProcessMapping}.

Mulgan~\cite{doi:10.1111/1467-9299.00218} describes an emerging view of responsibility as covering ``the internal functions of personal culpability, morality and professional ethics'', while accountability is ``concerned with the external function of scrutiny such as calling to account, requiring justifications and imposing sanctions''.

The notion of responsibility is commonly considered to have two distinct meanings. One meaning is the obligation of an agent to perform a specific task \cite{DBLP:conf/ecai/LimaRD10,DBLP:conf/deon/Mastop10}, known as \emph{task-based} \cite{Grossi2007} or \emph{forward-looking} \cite{DBLP:conf/ecai/LimaRD10,DBLP:conf/deon/Mastop10,DBLP:journals/see/DoornP12} responsibility.
The other meaning is that the agent's behaviour can be causally linked to the failure of a task. This is called \emph{causal}~\cite{Grossi2007} or \emph{backward-looking} responsibility~\cite{DBLP:conf/ecai/LimaRD10,DBLP:conf/deon/Mastop10,DBLP:journals/see/DoornP12}, which involves considerations of ability, knowledge and intentionality~\cite{DBLP:conf/ecai/LimaRD10,DBLP:conf/deon/Mastop10} and can be considered from a descriptive (who caused a failure) or a normative and evaluative (who should pay) stance~\cite{DBLP:journals/see/DoornP12}. However, even if an agent is deemed causally (backward-looking) responsible for a failure (or success), he/she may may not be considered blameworthy (or praiseworthy)~\cite{Shaver1985,Grossi2007,vandePoel2011}.
\label{sec:backward_responsibility}
Determining backward-looking responsibility becomes most challenging in the context of group activities, where it can be difficult or impossible to determine responsibility for a failure (the ``problem of many hands''~\cite{Thompson80_ManyHands}).




\subsection{Related work in the area of autonomous agents and multi-agent systems}
\label{sec:prior_mas_work}


Chopra and Singh \cite{ChopraSingh:RELAW2014,DBLP:conf/www/ChopraS16} describe accountability as a normative concept in the context of socio-technical systems: ``accountability requirements describe how principals ought to act in each other's eyes, providing a basis for their mutual expectations'' \cite{ChopraSingh:RELAW2014}. 
While we agree that norms and commitments are important for accountability, we argue that these are insufficient to capture all of the concept's elements, and that techniques for reasoning about norms and commitments (c.f., \cite{dastani-et-al:2017,fornara-colombetti:2010}) are only part of what is needed to create accountability-aware systems.

Baldoni et al.~\cite{DBLP:conf/aiia/BaldoniBMMT16} propose the study of \emph{computational accountability}. They consider accountability to be an ethical value, and define accountability as ``the acknowledgment and assumption of responsibility for decisions and actions that an individual, or an organization, has towards another party''. They note that, implicitly, ``individuals are expected to account for their actions and decisions when put under examination''. The paper focuses on multi-agent systems that track the state of conditional social commitments using business artifacts, in order to ``coordinate their activities, e.g., through responsibility assignment, as well as to identify liabilities''. It is argued that the ``analysis of accountability can be accomplished by looking at commitment relationships''.

In later work, Baldoni et al.~\cite{Baldoni:PRIMA2017,baldoni-et-al:2017b} take the viewpoint of \emph{accountability as a mechanism}, summarised by Bovens et al.~\cite{PublicAccountability} as ``an institutional relation or arrangement in which an agent can be held to account by another agent or institution''. They consider how such an institutional mechanism can be provided by design in a multi-agent system (MAS), and seek to provide ``structures that allow assessing who is accountable without actually infringing on the individual and private nature of agents'' and to ``determine action impact or significance by identifying the amount of disruption
it causes in terms of other agents and/or work affected''~\cite{Baldoni:PRIMA2017}. To this end, they present five ``necessary-but-not-sufficient principles that an MAS system must exhibit in order to support accountability determination''~\cite{Baldoni:PRIMA2017}. These principles state that (i) agents should interact within the scope of an organisation, (ii) must join the organisation by taking on a role, (iii) must be aware of the powers associated with a role before adopting it, (iv) can be accountable only for goals they have explicitly accepted, and (v) may specify the resources they need to satisfy a goal (which may be provided, or not, at the organisation's discretion). The fifth principle is endowed with particular significance for accountability determination: ``Should an uniformed agent stipulate insufficient provisions for an impossible goal that is then accepted by an organization, that agent will be held accountable because by voicing its provisions, it declared an impossible goal possible''~\cite{baldoni-et-al:2017b}. Baldoni et al.~operationalise these principles as an accountability protocol to be followed when an agent joins an organisation. 

Baldoni et al.~\cite{BaldoniJAAMAS22} examine how accountability relationships can be used to support robustness within an organisation by enabling unexpected events to be detected, and for handlers to exist to recover from such perturbations. They provide a conceptual model in which an \emph{accountability agreement} specialises the concept of a norm by recording the accountor and accountee, the condition under which a request for an account is allowed, one or more templates for an account and an association with a task. They provide a formalism for expressing the condition and the object of the agreement. The latter may be a complex task, broken down into a workflow of atomic tasks. An account is modelled as a sequence of events. Given a set of (forward-looking) responsibilities that assign agents to atomic tasks, they define an \emph{accountability structure}: a tree that explains how each agent can maintain situational awareness of the process it is involved in and construct an account of a workflow by composing accounts for the subtasks (possibly received from other agents). They show formally how a workflow can be made robust against a perturbation by adding a recovery strategy to handle an account of the perturbation. A number of rules are defined to link the state of accountability agreements with the activation of agent goals. They also present an implementation of this approach that extends the JaCaMo agent platform. 

This work of Baldoni et al.\ provides a well-founded practical approach for implementing the information phase of accountability in the context of engineering multi-agent systems, where agents are designed to be somewhat homogeneous, have a common view of a task, and a one-shot account comprising a sequence of events can be assumed to be sufficient. 
In contrast, we seek a more general model of computational accountability, allowing heterogeneous agents (possibly including people) in open multi-agent systems to participate in accountability relationships.
We also envisage a need for an open-ended debating phase due to differences in knowledge, plans, norms, etc. The needs of accountability in open multi-agent systems manifest additional research challenges, which we consider in Section \ref{sec:challenges}.

Many researchers in multi-agent systems and deontic logic have addressed the problem of determining backward-looking responsibility for a state of affairs, especially in the context of group plans or norms~\cite{10.1007/978-3-540-25927-5_9,DBLP:conf/ecai/MicalizioTT04,Witteveen:2005:DSM:1082473.1082596,Grossi2007,deJonge2009,DBLP:conf/deon/Mastop10,DBLP:conf/ecai/LimaRD10,BullingDastaniCLIMA2013,DBLP:journals/jair/MicalizioT14,LoriniResponsibility2014,Aldewereld:2016:GNM:2952298.2882967,Alechina:2017:CRB:3091210.3091279,ParkerGL23,MuNajibOren25}.
As this has been and remains an active area of research, in this paper we focus on other research issues in engineering accountable agents, and do not attempt to summarise this body of work.

Dignum~\cite{VDignumResponsibleAI17} addresses the question of how AI systems can be designed responsibly to ensure they are ``sensitive to moral principles and human value [sic]''. She discusses three principles of responsible AI: accountability, responsibility and transparency (ART). Accountability is described as ``the need to explain and justify one's decisions and actions to its partners, users and others with whom the system interacts''. Her view on responsibility focuses on developers and organisations using AI \emph{taking responsibility} for their AI systems, which includes ``developing theories, methods and algorithms to integrate societal, legal and moral values into technological developments in AI, at all stages of development''. 
%
%
Dignum~\cite{VDignumResponsibleAI17} defines \emph{transparency} as ``the need to describe, inspect and reproduce the mechanisms through which AI systems make decisions and learn to adapt to their environment, and to the governance of the data used or created''. 

Winikoff~\cite{10.1007/978-3-319-91899-0_1} considers the question of the \emph{trustability} of autonomous systems, i.e., how humans can come to trust them, and proposes three prerequisites for such trust: there should be a social framework for recourse; if the system makes a decision with negative consequences for the user, the system should be able to explain its behaviour; and the system should be subject to verification and validation to give assurance that key behavioural properties hold.

\subsection{Summary}\label{sec:summary}

\begin{figure*}[tb]
\includegraphics[width=\textwidth]{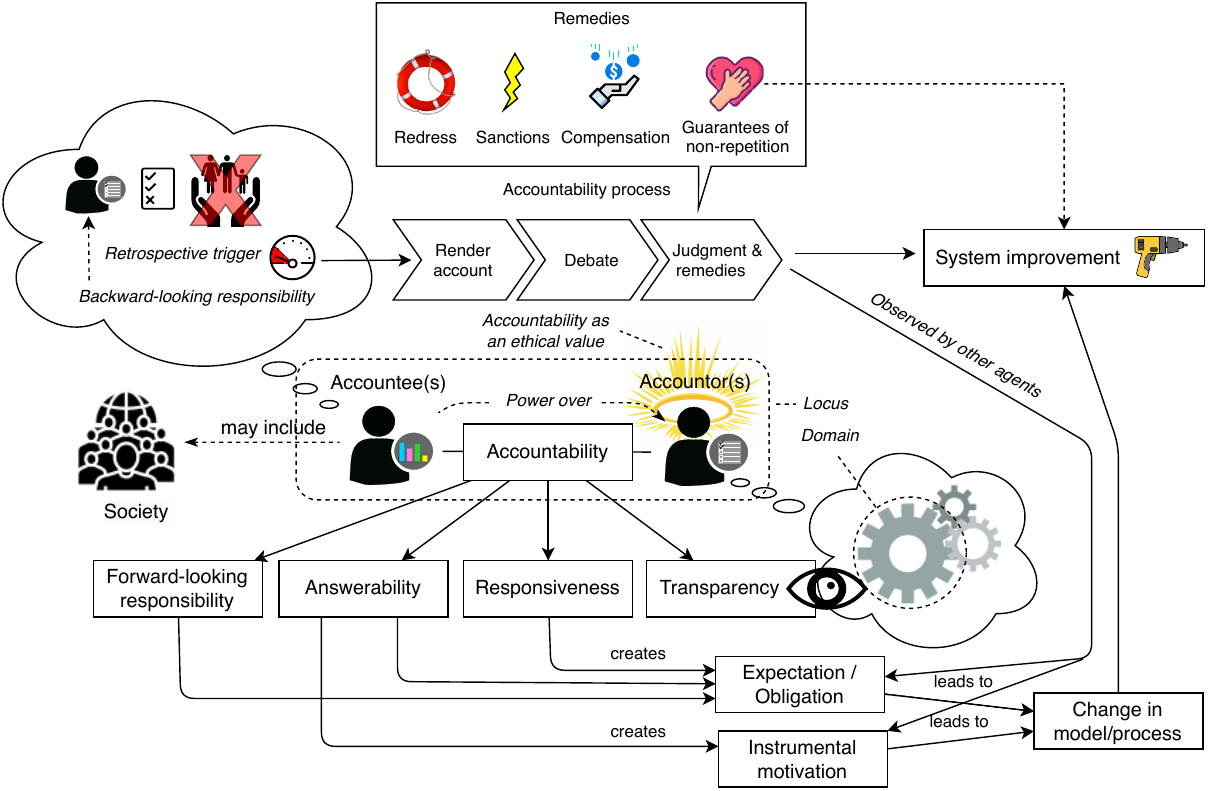}
\caption{Overview of accountability concepts from the literature}
\label{fig:accountability_concepts}
\end{figure*}

Figure~\ref{fig:accountability_concepts} presents a graphical summary of some of the key concepts related to accountability we have found in the literature.  The middle of the figure shows that accountability is a relational concept between an accountee (or account taker) and one or more accountors (or account givers). This could be a one-to-many, (less commonly) a many-to-one, or (rarely) a many-to-many relationship. The relationship implies the existence of a power relationship, in which the accountee can make demands of the accountor within a restricted domain (activities, practices or issues \cite{AccountabilityHealthCare}).

While an accountor can be made accountable in order to facilitate system improvement in various ways, we also recognise that accountability fosters ethical behaviour \cite{dubnickAccountabilityEthicsReconsidering2003}, and can be viewed as a desirable value in and of itself.


At the top left of the figure, we see that accountability is triggered by the accountee becoming aware of a substandard performance or failure of a task or service performance for which the accountor is considered responsible. This occurs post-event, and therefore reflects backward-looking responsibility. This trigger initiates a three-phase process in which the accountor is required to render an account, an interactive debate (or an interactive and iterative elaboration of the account) is performed, followed by rendering of a judgment and (possibly) a required remedy. The icons shown at the top of the figure denote different types of remedy (adopting the categories of Gibbons \cite{gibbonssocialaccountability2014} summarised in Section~\ref{sec:remedies}).

The lower part of the figure illustrates some of the concepts related to and/or required for accountability. First, the accountor must understand their delegated \emph{responsibility} for specific tasks and/or goals---this is forward-looking responsibility (in the sense that they should plan their actions taking this understanding into account). Second, \emph{answerability} is required: the accountee must provide an account (potentially iteratively) as required by the accountee. Third, we highlight \emph{responsiveness} in the sense of Koppell (Section \ref{sec:koppell_responsiveness}), but generalised beyond the field of public administration: the accountor should be proactively responsive to the demands and goals of the accountee(s). Finally, there must be \emph{transparency}: proactive or demand-driven dissemination of information about the accountor's activities and performance \cite{FoxTransparencyAndAccountability}.

On the upper right, the figure highlights that a key outcome of accountability is system improvement. This may be a direct outcome of the accountability process as an imposed remedy. Alternatively, this may be due to an agent improving its model and/or process to avoid the costs of possibly being held to account in the future
or to ensure its obligations as an accountor are met. These motivations and obligations may come not only \emph{directly} through the accountability relationship between the accountor and accountee, but also \emph{indirectly} by observing the outcomes of accountability processes involving other accountors. We also consider that obligations are created by forward-looking responsibility (to perform the delegated task) and responsiveness (to attend to the demands and needs of the accountor and other parties being served). We consider that transparency is more difficult to encapsulate by norms due to this concept's abstract nature, and believe this may be better thought of as a design goal for the accountor.

\section{Example application of accountability in a multi-agent system}
\label{sec:scenario}

In this section, we present a fictional scenario of the accountability process at work in a multi-agent system comprising  of people and software agents. This example illustrates how some of the key features of accountability discussed in the previous section can produce improvements to a socio-technical system in which software agents have the capability to participate in accountability relationships.

Consider an earthquake relief scenario where ground robots and aerial drones sent from a range of countries are given specific roles. There is a set of drones ($M_1, M_2, \ldots$) responsible for aerial mapping and identifying safe zones, a single large aerial delivery drone $A_1$, and three sets of ground robots for supply delivery ($S_1, S_2, \ldots$), debris clearance ($D_1, D_2, \ldots$), and casualty rescue ($C_1, C_2, \ldots$). They are supervised by a human operator, who sets mission-level priorities and norms, overrides agent decisions in critical situations and holds the robots to account when things go wrong with the aim of preventing reoccurrences.

Suppose that robot $S_1$ is assigned to deliver supplies to zone $X$ while $C_3$ is assigned to rescue casualties in the same area. Now assume that $S_1$'s operations block $C_3$ (e.g., due to narrow routes). Since casualty rescue is deemed to be more important, $S_1$ agrees to reroute, delaying delivery by an hour.

However, while rerouting, $S_1$ encounters an uncharted hazard, requiring $D_8$ to be retasked and delaying delivery by several more hours, causing the condition of wounded in the zone of interest to significantly worsen.

This is noticed by the operator who requests an account from the system. $S_1$ starts by providing its path plan and the operator asks why a more direct route was not chosen, with $S_1$ responding that $C_3$ was in the way. The operator then asks $C_3$ why it was there and it responds that it was undertaking a high priority rescue. The operator accepts this and queries $S_1$ as to why it was delayed, discovering that the unmapped obstacle delayed it. The operator asks why $S_1$ did not take the option available to it of requesting a special aerial mapping mission. $S_1$ replies that it has a norm stating that this should only be done for missions with priority greater than 5 (which was the case for its mission) if its latest map information is a day or more out of date (which was not the case: it was only two hours out of date).


The operator decides that all robots should modify this norm: in an earthquake scenario the map staleness threshold can be lowered to one hour. Furthermore, concerned by the unmapped obstacle, the operator instructs all mapping drones to increase their mapping frequency for the next week by postponing maintenance checks.
The operator also orders the remedy of sending $A_1$ to deliver the urgently needed supplies to the wounded  citizens.

The next day, when temperatures have fallen dramatically, an aerial mapping drone's battery rapidly loses charge during a mapping mission and the drone falls out of the sky. The operator's supervisor queries this (as part of an accountability process involving the operator and the drone in question) and discovers that
the drone's pre-mission battery check did not take temperature into account. This omission is likely due to the drone having been donated from a country with a warm climate. The operator sends all drones a battery discharge curve parameterised by temperature and requires all drones to consider this during their pre-mission battery check.

A few days later, several mapping drones stop functioning. An accountability process initiated by the supervisor determines that this is due to the reduced maintenance schedule that allowed the higher frequency mapping to occur. The operator who ordered that is found to have failed to consult a maintenance technician about the likely impacts. A new norm is put in place to require this consultation before maintenance is reduced. In making this change, and that from the previous paragraph, the supervisor has considered the trade-off between the remedies made and the overhead they create (although it could be argued that only drones from countries with warmer climates needed a change to their battery check process).

\section{Some Research Challenges in Developing Accountability-Aware Agents}
\label{sec:challenges}


As noted in Section~\ref{sec:prior_mas_work}, accountability has been used to drive desirable behaviour within a MAS by, for example, underpinning interaction protocols \cite{baldoniRobustnessBasedAccountability2021}.
In contrast, we consider open multi-agent systems where such protocols may not be specified ahead of time, agents may be under the control of different organisations, and the agents may have different goals. In this section we discuss the degree to which accountability in such a context raises research problems that go beyond straightforward computational encodings.

Accountability can affect the agent at multiple points during interaction, namely before, during, and following task execution. Before task execution, the agent may need to plan how to undertake a task, recognising that they may be held to account. This could involve, for example, selecting plans that include logging or are recognised as best practice. During task execution the agent may need to decide what information to store in case they are held to account. Then, following task execution, 
%
%
those capable of asking for an account must decide whether and when to do so, triggering the start of the accountability process. 

In the next phase, accountors and accountees request and exchange information. Following this, a judgment must be made as to whether the account-taker acted appropriately, and any remedies that are needed must be put into place.

In the remainder of this section we focus on challenges around the computational aspects affecting each of these points in the process.



\subsection{Optimisation vs.~satisficing}
The accountability process is triggered by the detection of the failure or poor performance of a task or some undesirable outcome. This is not the same thing as a \emph{suboptimal} outcome. For example, if a taxi driver has estimated that our trip to the airport will take between 20 and 30 minutes, we can't reasonably hold them to account if the trip takes 25 minutes rather than the optimal time (within the estimated range) of 20 minutes. In a more formal context, service-level agreements are commonly expressed in terms of minimal levels of required performance rather than the theoretically best performance the service provider might be capable of. This suggests that underlying models used when reasoning about task performance subject to accountability and the trigger for initiating an accountability process should be based on \emph{satisficing} rather than optimisation or a combination of both as in Goal Programming \cite{PracticalGoalProgrammingChap1}. This reflects the usual lack of complete information and the uncertainty of outcomes in complex real-world scenarios as well as Simon's view \cite{simon_models_of_man} that human rationality is inherently bounded by limited cognitive capabilities and time available for decision-making. While software agents may be able to make optimal decisions for a larger range of problems than people, in socio-technical systems the bounded rationality of humans will still be present.


The use of satisficing in multi-agent decision-making remains underexplored, and accountability introduces distinctive challenges that go beyond simply identifying a performance threshold. Where a task involves multiple criteria — timeliness, resource use, safety, quality of outcome — an accountee must determine how to aggregate these into a coherent standard, and how to weight them when they conflict. This is further complicated when the accountor and accountee operate under different informational assumptions: the accountee's threshold may be calibrated to what they believe is achievable given their model of the environment, while the accountor must plan under the actual constraints it faces. A deeper challenge arises when satisficing and optimising goals coexist within the same agent: an agent may be required merely to satisfy a service-level obligation in one dimension while simultaneously optimising another, and accountability processes must be sensitive to this distinction when judging whether performance was adequate. Finally, satisficing thresholds are not static — they may need to evolve as the capabilities of accountors become better understood, as environmental conditions change, or as prior accountability processes reveal that existing standards were poorly calibrated. How an accountee should revise thresholds in a principled way, without either systematically under-demanding or imposing unachievable standards, is a research question that connects accountability directly to problems of learning and norm revision in multi-agent systems.

\subsection{Accountability with respect to best practice in an uncertain world}
The real world is complex and unpredictable (except at a coarse grain). Any task could fail due to unanticipated or unusual events. The question immediately arises as to how a computational model of accountability can take this into account without allowing accountees an easy get-out-of-jail card by blaming random chance. Human institutions handle this routinely by deciding on levels of acceptable risk and appealing to notions of best practice. For example, a city council might decide to maintain their storm water drains to handle a 1 in 20 year deluge of rain but not a 1 in 100 year event. This is more likely to be defensible after widespread damage occurs due to an extreme flood if it can be argued that this is best practice nationally or internationally. This raises two questions about formalising and operationalising the concept of best practice:
\begin{itemize}
\item How can best practice be modelled? Is it a set of norms, or is more needed such as justifications or evidence? And is the articulation of best practice a one-shot event, or should it have an explanatory, dialogical or argumentative nature?
\item What are the dynamics of best practice? That is, how does best practice become established, how does it change over time and how do (visible) accountability processes effect changes to best practice?
\end{itemize}

\subsection{What information must be kept to (potentially) render an account?}
The answerability requirement of accountability means that an accountable agent must not only be willing to render an account to an accountee, but it must also be prepared for this eventuality. In particular, it must ensure that sufficiently complete records of the information that is relevant to the performance of each task are kept, including the relevant world state at decision points, the options considered and taken (with justifications) relative to whatever planning or process model it uses, and the observed results. We will refer to this as the \emph{trace} of the task, to distinguish it from other information that is not specific to a given task performance, such as the general method it uses to make its decisions (e.g.\ a specific planning method) and the generic inputs to that process, such as the agent's plan library (although specific plans may be included in an agent's account).

The questions of what makes the elements of a trace \emph{relevant} and the entire trace \emph{sufficiently complete} are complex and the answers may change over time. The accountee's and accountor's notions of relevance may not coincide, it may not be clear what granularity of information should be recorded, and events out of the agent's control may change relevant world state--should the agent be monitoring those with some required frequency? Furthermore, there may be privacy issues if the accountor is working on multiple concurrent tasks involving different involved parties. In reality, there may be no provably correct choice and the requirements for the trace may be the subject of organisational norms.

\subsection{When should an accountability process be initiated?}
As illustrated in Figure~\ref{fig:accountability_concepts}, the accountability process is initiated by the detection of an undesirable outcome of some task performance. This is likely to be domain-specific and therefore we do not associate any research questions to  this task. 

The next step is to determine which party or parties have (backward-looking) responsibility for this outcome. As noted in Section~\ref{sec:backward_responsibility}, this problem has been well studied in the multi-agent systems community and we have not identified any new aspects of this problem that are specific to accountability.

The accountee must then decide whether it is worthwhile to incur the cost of undertaking an accountability process or whether they are required to do so by organisational or social norms. For example, in the taxi scenario described above, assume that the taxi arrives at the airport 31 minutes late. A decision must be made as to whether to hold the taxi driver to account. In deciding  whether to do so there is a need to balance the expected benefits of doing so with the costs. These benefits could accrue to the individual agent or to the wider society, and characterising such reasoning is challenging. Costs in holding someone to account revolve around the time and effort in doing so, but may also include structural costs (e.g., in human societies, the time and monetary cost of an ombudsman, enquiry or court proceedings). In contrast, calculating the system-wide benefits such as protecting the public (or organisation) from future occurrences or improving professional standards can only be predicted based on (often speculative) models or heuristics (see Section~\ref{sec:future_benefits}).

The cost/benefit analysis is also highly context dependent. In the taxi example, being 31 minutes late and thus missing an expensive flight may mean that holding the driver to account will result in some compensation, but in many scenarios, the costs can outweigh the benefits.

\subsection{What counts as an account?}
When the accountor is asked to give an account of its performance of a task, how can it decide what information to provide? One option is to send the entire trace, and possibly also the decision-making mechanism it used (e.g., its planning method), and leave it to the accountee to analyse this information. This assumes that the accountee has the capability, resources and attention available to perform such a comprehensive analysis. If the accountee is human, this assumption is highly unlikely to hold. An alternative is to place more responsibility on the accountor to choose a higher-level abstraction of the trace to give a focused answer to the accountee, which the accountor and accountee can iteratively refine during the subsequent debating phase.
Identifying an appropriate abstraction of this trace is an open research question which must consider the accountee's capabilities and informational needs as well as the bounds or scope of the query, as many elements of the accountor's trace may not be relevant.

An earlier version of this paper \cite{CranOrenVasc_2019} presented a semi-formal analysis of key components of the notion of answerability in the context of accountability. This defined the concept of a \emph{valid reply} to a specific accountability query from an accountee to an accountor, given a particular \emph{scope} of inquiry (a generalisation of the notion of domain) within this accountability relationship and covering a specific time period. This analysis is omitted in this viewpoint paper. 

\subsection{Defining a protocol for accountability dialogues}

Within the debating phase, the accountee seeks to obtain sufficient information from the accountor, and from any other stakeholders, to reach a judgement. Depending on the context, this phase may be relatively informal (e.g., a boss asking an employee what happened), or formal (e.g., a tribunal meeting with minutes, etc.). The dialogue may involve requests for further information from the accountor, `why' questions to help with debugging and `what-if' questions that explore counterfactual modifications of the accountor's domain knowledge, task model, reasoning mechanism, norms, etc., with a view to improving outcomes in the future.

When considering computational agents, there is a need for a protocol to drive this phase. Research into \emph{dialogue games} \cite{walton1995commitment} as well as agent communication languages \cite{poslad07specifying} has identified how speech acts undertaken by agents can affect each other's mental state, and this can underpin the debating phase. The dialogue used here can be very complex and include different elements of multiple types of dialogue (inquiry, information seeking, persuasion, etc.) and challenges include both dialogue specification and dialogue strategy in this context.

\subsection{Rendering judgment and selecting remedies}

Once the accountee has gathered sufficient information they are able to render judgement and, if necessary, select one or more remedies. This process proceeds in two broad phases; first, a decision must be reached as to whether the accountor is liable for the poor task performance. If so, the accountee may choose to apply an appropriate sanction. This decision may be expected to conform to principles that promote the system-wide value of accountability, e.g., protecting the public (or organisation) and setting professional standards \cite{Roberts_v_PCC_NCNZ}. Therefore, prior judgments must also be an input into the judgment phase. 

Even if the accountor is not found liable, the accountee may still need to select and apply one or more remedies. For example, if a contractor demonstrably followed all current health and safety regulations, they may not have a case to answer for, but a tribunal may recommend that additional health and safety processes be put into place to prevent an accident from reoccurring.

A key problem that occurs in human organisations is the imposition of increasing layers of bureaucracy, often as an outcome of past accountability processes. For example, it seems a common refrain from academic staff of universities that their universities do not trust them to undertake university processes competently, so they must provide increasing amounts of documentation for approval. This will also be a risk when software agents act as accountees. It is an important research challenge to find effective ways to target ``system improvement'' remedies to the agents that have proved to need them, without adding extra overhead to processes more widely. A trade-off must be made between the risk of future repetitions of a problem and the increased costs that a remedy may cause. 

The challenge of introducing appropriate sanctions or modifying the system to achieve some outcome revolves around appropriate incentivisation of agents and areas including mechanism design, norm synthesis \cite{shoham95social,morales15online,AnavankotNormSythesis} and game theory can provide insight to this problem.

\subsection{Modelling the future benefits of accountability}
\label{sec:future_benefits}
Several phases of the accountability process outlined above involve making choices to incur costs informed by predictions of future positive outcomes, e.g., when an accountor maintains task execution traces in case of future accountability queries, an accountee choses to initiate an accountability process and an accountee chooses a sanction and/or remedy. Techniques such as reinforcement learning are widely used to reason about future benefits in the context of uncertainty, and integrating accountability related actions into such a framework poses another future research challenge.


\subsection{Reactive and proactive responsiveness}
Koppell's fifth dimension of accountability---responsiveness (not to be confused with answerability; see Section~\ref{sec:related_concepts})---revolves around providing attention to the demands and needs of the people being served. Even outside Koppell's specific concern of public administration, accountability is often related not to a specific well-defined task but rather to a higher level maintenance goal \cite{duff14maintenance} (corresponding to Koppell's notion of a \emph{need}). For example, a robot tasked with keeping a communal office kitchen clean and stocked with provisions should clean the office more often when a major project deadline is looming. It could be reactive to the more frequent disorder of the kitchen or it might proactively request a second robot to help it after predicting the increased disorder.

Koppell also refers to the elicitation of the served community's \emph{demands}. In an open system of heterogeneous agents (possibly including people), dynamic elicitation involves learning new actions or states of the world that may be desired. In some cases, the agent may have the capability and resources to satisfy these needs but lack the necessary procedural knowledge. In this case, techniques from planning \cite{nau04automated} and robot learning \cite{gu17deep} could be applied. Where direct elicitation is not feasible, if there is a measurable level of satisfaction for its services, it could apply a reinforcement learning approach to explore alternative or enhanced service activities (perhaps advised by a large language model \cite{ryu2025curricullm}). For example, an agent that provides news articles relevant to users' interests might find that the proportion of users accessing these articles is reducing. It might then try adding links to fact-checking web sites, but later abandon this when those links were seldom accessed.


\subsection{Formal modelling of accountability-related norms}

Section~\ref{sec:what_is_accountability} discussed a range of accountability-related concepts, many of which can be considered as norms. Some are intrinsic to the accountability relationship itself, e.g., answerability, forwards-looking responsibility for a task, responsiveness and keeping appropriate logs, while others are specific to the application domain, e.g., organisational best practice and norms of sanction and remedy selection. Elements of accountability are underpinned by communication. Therefore, in the context of open multi-agent systems, some form of ontology to encode these and other accountability-related concepts (such as accountability norms) would potentially be required, as would a description of their structure and lifecycle.

\subsection{LLMs as accountability advisers}
As Section~\ref{sec:what_is_accountability} and the example scenario in Section~\ref{sec:scenario} have shown, accountability is a complex multi-faceted concept and in complex domains, the debate, judgment and remedy phases may be more akin to creative activities than searches for optimal solutions. Large language models (LLMs) may provide a valuable tool to assist with these phases due to their ability to generalise from their vast amounts of textual training data, in which untold examples of accountability will have appeared. Evaluating the use of LLMs for a range of accountability-related tasks and the reduction and mitigation of hallucinations in this area are important research topics.

\subsection{Accountability testbeds and competitions}

Across MAS and AI research, the establishment of testbeds and competitions has helped to build communities of researchers with a coherent focus \cite{trading01,ipc20,icaps23,mapc22,iccma23}. We believe this would also be true for computational accountability. Developing an accountability testbed would be a valuable contribution, but raises several significant challenges. Below, we sketch out some ideas for such a testbed.
Developing a testbed would involve simulating a complex multi-agent task, such as the earthquake rescue scenario from Section~\ref{sec:scenario} and choosing specific acccountability-related competencies that could be tested in isolation while others were hard-coded or manually scripted. Elements that the testbed could then exercise include
\begin{itemize}
  \item
    Account soundness: determining when an account provides a sound explanation of the observed behaviour (completeness may not be possible in a dynamic stochastic environment).
  \item
    Model alignment: determining when either the accountee or accountor can conclude that the other's model of the domain does not align with its own, based on the account given. An additional challenge would involve determining these differences.
  \item
    Sanction and remedy selection: modelling and reasoning about the expected effects of different options for sanctions and remedies, and selecting one or more to apply. These effects may be both direct impacts of sanctions and remedies applied directly to the accountor as well as those from wider society-level signalling and more widely applied remedies.
    The approach could be based, for example, on a learned reward model, predefined values, or both.
  \item
    Accountability initiation: deciding when an accountee should initiate an accountability process. 
\end{itemize}


Once solutions to these specific problem (and others that emerge as key components of accountability) are developed, the testbed could be used to evaluate the entire accountability process taking place within a fixed agent organisation over a period of time. One way to achieve this might be to use a dynamic, stochastic and partially observable environment that is initially relatively predictable and involves tasks that require limited coordination between agents. Agents playing different roles could be provided with plans and norms that are sufficient to achieve the tasks in the initial environment. The environment could then be made gradually more complex (especially in terms of the need for coordination) while organisational level goals, values and norms also change periodically. The challenge would then be to use accountability to achieve system improvement or to maintain system performance as the environment and organisational goals change.

\subsection{Summary}

To instantiate accountability-aware agents capable of acting as accountors and/or accountees requires integrating multiple technologies, primarily centred on reasoning and communication.
\begin{itemize}
  \item Satisficing-based planning, goal selection and progression drive the accountor's planning cycle, and are used to decide how agents should act in pursuit of their goals. This element must be adapted to the complexity of the domain, for example incorporating elements such as reasoning about uncertainty.  While there have been computational approaches for reasoning about satisficing goals in the fields of operations research and management research \cite{PracticalGoalProgramming,SimEtAl_RobustSatisficing}, there has been limited attention to satisficing in multi-agent systems \cite{ijcai2023p32}. 
  
  \item 
  Norms govern and influence various aspects of reasoning about accountability for both the accountor and accountee. For example, norms affect
  how the accountor's task should be performed, what information should be stored in the accountor's trace, what activities are considered within the scope of the accountability relationship, whether the accountor can be considered liable for a failure or poor outcome and what sanctions and remedies are appropriate.
  The accountee's judgement may establish new norms governing the accountor's behaviour, or the behaviour of the agents in the system more generally.
  
  The role of norms in multi-agent systems has been studied for more than 25 years \cite{Shoham1997,DBLP:journals/ail/ConteFS99,Conte2001,DBLP:journals/cmot/BoellaTV06,Normas09SpecialIssue,NormativeAgentsInATBook2013,DBLP:conf/dagstuhl/2013dfu4,HandbookOfNorMAS}, and there are many existing representations for norms and approaches to integrating them into agent reasoning that could be applied in the context of accountability.


  \item Dialogical protocols (c.f., dialogue games \cite{walton1995commitment}) as well as work on agent communication languages \cite{labrou99agent} underpin the debate phase. Given the complexity of such dialogues, reasoning about utterances (c.f., dialogue strategy \cite{Rienstra:2013,snaith2020argument}) must be examined. Research on human-agent dialogues to debug the agent's belief-desire-intention (BDI) agent program using `why' questions \cite{DBLP:conf/atal/Winikoff17,DennisOrenBDIExplanationDialogue} provides a good starting point, although we believe that counterfactual questions are also necessary.
  
  \item Various forms of reasoning are required to identify interventions, sanctions and reparations, as well as determine whether to initiate the accountability process. This reasoning will potentially incorporate game theoretic, principal agent theoretic and normative elements. Since utility underpins many such types of reasoning, examining how accountability shapes and affects utilities, so as to provide appropriate incentives and rewards, is also necessary.
\end{itemize}






\section{Conclusions} \label{sec:conclusions}
This paper issues a call to arms for the multi-agent systems (MAS) community: to move beyond engineered compliance and toward the development of agents that can meaningfully participate in accountability processes. We envision agents that act as accountors and/or as accountees. In the former case, such agents can reason about what information they should store and participate in discussions to demonstrate how, and why, they have acted appropriately. As accountees, such agents should be able to decide whether to initiate an inquiry into the actions of others; probe accountees to understand why they acted as they did; pass judgment as to the appropriateness of action; and utilise their powers to change and improve system behaviour.

In human societies, accountability fosters trust, supports learning, and enhances system robustness in the face of failure. We argue that these same benefits can and should be realised in agent societies, particularly as autonomous systems become increasingly embedded in socio-technical contexts.

While prior MAS research has acknowledged the importance of accountability, the focus of such work was often on closed systems and software engineering perspectives. In contrast, we advocate for a richer, more nuanced treatment of accountability---one that reflects its socio-technical complexity and draws from interdisciplinary insights.

To this end, we have outlined a research agenda that spans multiple forms of reasoning ---satisficing, normative, practical, game-theoretic, and decision-theoretic --- across the full accountability lifecycle. We have also highlighted the central role of structured dialogue in enabling agents to explain, justify, and contest actions and outcomes.

Ultimately, accountability is not only a mechanism for ensuring desirable behaviour; it is a catalyst for advancing foundational research in agent reasoning, communication, and interaction. By embracing this challenge, the MAS community can help shape autonomous systems that are not only intelligent, but also trustworthy, transparent, and socially aligned.

\bibliographystyle{unsrt}
\bibliography{accountability}


\end{document}